\documentstyle[psfig]{article}

\newcommand{\calN}[1]{#1}

\begin{document}

\title  { Analyzing X-ray variability by Linear State Space Models }

\author {       Michael K\"onig \\
             Institut f\"ur Astronomie und Astrophysik,\\
	       - Astronomie -\\
	       Universit\"at T\"ubingen,
	       Waldh\"auser Str. 64,
	       D -- 72076 T\"ubingen
 \and
		Jens Timmer\\
	       Fakult\"at f\"ur Physik,
	       Albert-Ludwigs-Universit\"at \\ 
	       Hermann-Herder Str. 3,
	       D -- 79104 Freiburg\\
               Freiburger Zentrum f\"ur Datenanalyse und Modellbildung,\\
               Albert Str. 26-28,
               D -- 79104 Freiburg	}

\maketitle

\begin{abstract}
In recent years, autoregressive models have had a profound impact on the
description of astronomical time series as the observation of a stochastic
process. These methods have advantages compared with common Fourier
techniques concerning their inherent stationarity and physical
background. However, if autoregressive models are used, it has to be taken
into account that real data always contain observational noise often
obscuring the intrinsic time series of the object. We apply the technique
of a Linear State Space Model which explicitly models the noise of
astronomical data and allows to estimate the hidden autoregressive
process. As an example, we have analysed the X-ray flux variability of the
Active Galaxy NGC~5506 observed with EXOSAT.
\end{abstract}
\section {Introduction}

A common phenomenon of Active Galactic Nuclei, which presumably harbor
supermassive black holes with masses of $10^{6}$ -- $10^{9}\,M_{sol}$
(Rees 1984), is the strong variability which can be observed in 
X-ray lightcurves. These AGN lightcurves seem to show
featureless 'red noise', i.e. scale-free, divergent variability at
low frequencies, often also described as flickering or $1/f$ fluctuation
(Lawrence et al. 1987). The $1/f$ term describes the power law
distribution of the spectral power with the function $f^{-\alpha}$ in
the power spectrum, often denoted as '$1/f$' behavior.

We present an alternative model to analyse the variability seen in
the X-ray lightcurves of AGN. The standard method of analyzing time
series in the frequency domain is discussed briefly in Section 2. The
alternative is known as a Linear State Space Model (LSSM) based on the
theory of autoregressive processes (Scargle 1981, Honerkamp 1993)
which usually cannot be observed directly since the observational
noise (i.e. detectors, particle background) overlays the process
powering the AGN. A LSSM fit applied to the time series data yields
the dynamical parameters of the underlying stochastic process.  These
parameters should be strongly correlated to the physical properties of
the emission process. The corresponding LSSM power spectrum exhibits
both the decrease of power at medium frequencies and a limitation of
spectral power at low frequencies. The detailed mathematical
background of LSSM and the fit procedure are described in Section~3
and~4. Finally we present first results using this technique with EXOSAT
data from the Seyfert galaxy NGC~5506 in Section~5.

\section {Description of the Method based on the $1/f^{\alpha}$-model}

Although measured astronomical data are time domain data, a commonly
applied method works in the frequency domain by analyzing the power
spectrum of the time series. As the observational window function is
convoluted to the true spectrum of the source, artefacts might be produced
in the power spectrum, which make a proper interpretation more difficult
(Papadakis and Lawrence 1995, Priestley 1992). In most cases, the power
spectra are fit by a power law function with an offset described as
$1/f^{\alpha} + c$, with values of $\alpha$ ranging from 0 to 2 and a mean
of about 1.5 (Lawrence and Papadakis 1993). The value $c$ is often denoted
as the `observational noise floor' which describes the random process
comprising the observational errors whereas the `red noise' component is
the signal of interest. In the case of long AGN observations, however, a
flattening at low frequencies occurs which cannot be modelled by the
$1/f^{\alpha}$-model (McHardy 1988).

The $1/f^{\alpha}$-model is an ad hoc description of the measured
periodogram, without any direct physical motivation. However, it is
possible to generate time series with a $1/f^{\alpha}$-spectrum using
self-organized criticality models simulating the mass flow within an
accretion disc of the AGN (Mineshige et al. 1994). Such models produce a
stationary time series that exhibits a $1/f^{\alpha}$-power spectrum by
limiting the timescales occurring in the simulated accretion process. A
$1/f^{\alpha}$-model without limited timescales would be stationary only if
the power law slope is smaller than unity (Samorodnitsky and Taqqu
1994). The observed time series is composed by the superposition of single
luminosity bursts. The slope of the $1/f^{\alpha}$-spectrum of data
simulated in that way is about 1.8, significantly higher than those
measured from real data (Lawrence and Papadakis 1993). If the inclination
of the accretion disk is brought in as an additional model parameter the
slope can be diminished, but not in a way that leads to convincing results
(Abramowicz et al. 1995). Another point that contradicts this assumption
is that there is no correlation between the spectral slope and the type of
the Seyfert galaxy (Green et al. 1993). This correlation should be present
since the Seyfert type is believed to be caused by the inclination of the
line of sight (Netzer 1990).

The periodogram which is used to estimate the true source spectrum is
difficult to interpret in the presence of non-equispaced sampling time
series arising from real astronomical data (Deeter and Boynton 1982 and
references therein). The estimation of the $1/f^{\alpha}$-spectrum is
hampered even in the absence of data gaps. This is due to the finite extent
of the observed time series. Therefore, the transfer function (Fourier
transform of the sampling function) is a sinc-function which will only
recover the true spectrum if this is sufficiently flat (Deeter and Boynton
1982; Deeter 1984). In the case of `red noise' spectra the sidebands of the
transfer function will cause a spectal leakage to higher frequencies which
will cause the spectra to appear less steep (the spectral slope will be
underestimated).

Even periodograms of white noise time series deviate from a perfectly flat
distribution of frequencies as the periodogram is a
$\chi^{2}_{2}$-distibuted random variable with a standard deviation equal
to the mean (Leahy et al. 1983). Thus the periodograms fluctuate and
their variances are independent of the number of data points in the time
series. Due to the logarithmic frequency binning, AGN periodograms will always
show this strong fluctuation due to the low number of periodogram points
averaged in the lowest frequency bins (see Fig.1).

\begin{figure}
{Fig.1: a) EXOSAT ME X-ray lightcurve of the quasar 3C273 (Jan.~1986), 
b) corresponding periodogram. Each dot represents the spectral power at its
frequency, stepped with $1/T_{\rm{tot}}$. The periodogram is binned
logarithmically (squares indicates a single point within the frequency
bin). }
\end{figure}

Furthermore, additional modulations can be created in white noise
periodograms if the time series consists of parts which slightly differ in
their means and variances, respectively (Krolik 1992). In the case of the
EXOSAT ME X-ray lightcurves this effect is due to the swapping of
detectors as each detector has its own statistical characteristics which
cannot be totally suppressed (Grandi et al. 1992; Tagliaferri et
al. 1996). Fig.1a shows a typical X-ray lightcurve which mainly consists of
uninterrupted 11 ksec observation blocks before detectors are swapped. If
the periodogram frequency corresponds to the observation block length, the
calculated sum of Fourier coefficients equals its expected white noise
value of $\sigma^{2}$ due to the constant mean and variance within the
entire oscillation cycle. At other, mainly lower, frequencies the Fourier
sum yields non-white values due to temporal correlations caused by
different means and variances of observation blocks located in the test
frequency cycle. These deviations from a flat spectrum will be very strong
at frequencies which correspond to twice the observation block length. The
arrows in Fig.1b clearly show this minimum feature at $9.1\cdot 10^{-5}$Hz
and another shortage of power at $1.4\cdot 10^{-5}$Hz which corresponds to
the long uninterrupted 72 ksec observation block starting at the second
half of the EXOSAT observation (Fig.1a).

Consequently a model is required which operates in the time domain and
avoids any misleading systematical effects occuring in power spectra.

\section {Mathematical Background of the Linear State Space Model}

In this section we briefly introduce the Linear State Space Model
(LSSM). For a detailed discussion, see Honerkamp (1993) and Hamilton
(1995). The LSSM is a generalization of the autoregressive (AR) model
invented by Yule (1927) to model the variability of Wolf's sunspot
numbers.

We follow Wold's decomposition theorem (Wold 1938; Priestley 1992; Fuller
1996) which states that any discrete stationary process can be expressed as
the sum of two processes uncorrelated with one another, one purely
deterministic (i.e. a process that can be forecasted exactly such as a
strictly period oscillation) and one purely indeterministic. Further, the
indeterministic component, which is essentially the stochastic part, can be
written as a linear combination of an innovation process, which is a
sequence of uncorrelated random variables.

A given discrete time series $x(t)$ is considered as a sequence of
correlated random variables. The AR model expresses the temporal
correlations of the time series in terms of a linear function of its past
values plus a noise term and is closely related to the differential
equation describing the dynamics of the system. The fact that $x(t)$ has a
regression on its own past terms gives rise to the terminology
`autoregressive process' (for detailed discussions see Scargle 1981;
Priestley 1992). A time series is thus a realization of the stochastic
process or, more precisely, the observation of a realization of the process
during a finite time interval. The AR model expresses the temporal
correlations in the process in terms of memory, in the sense that a filter
($a_i$) remembers, for a while at least, the previous values $x(t-i)$. Thus
the influence of a predecessor value decreases as time increases.  This
fading memory is expressed in the exponential decay of the AR
autocorrelation function (see eq.~\ref{eq:rar}). The AR processes variable
$x(t)$ remembers its own behavior at previous times, expressed in a linear
relationship in terms of $x(t-1), x(t-2),\ldots$ plus $\epsilon(t)$ which
stands for an uncorrelated (Gaussian) white noise process.
\begin{eqnarray}
    x(t) = \sum_{i=1}^p a_i x(t-i)  +  \epsilon(t),
       \qquad \epsilon(t) \sim {\calN{N}} (0,\sigma^2) \label{eq:ar}
\end{eqnarray}
The number of terms $p$ used for the regression of $x(t)$ determine the
order of the AR process, which is abbreviated to an AR[p] process.
The parameter values $a_i$ have to be restricted for the process to be
stationary (Honerkamp 1993). For a first order process this means $
|a_1| < 1 $, for a second order process: $\left|a_1 \pm
\sqrt{a_1^2+4a_2^2}\,\right| < 2$. Depending on the order $p$, the parameters
$a_i$ of the process represents damped oscillators, pure relaxators or
their superpositions. For the first order process AR[1] the relaxation
time $\tau$ of the system is determined from $a_1$ by:
\begin{eqnarray}
\tau = -\frac{1}{\log{|a_1|}} \label{eq:loga}
\end{eqnarray}
In the case of a damped oscillator for an AR[2] process the parameters,
the period $T$ and the relaxation time $\tau $ respectively, are
related by:
\begin{eqnarray}
  a_1 & = & 2 \cos \left(\frac{2\pi}{T}\right) \, e^{-1/{\tau}} \\ 
  a_2 & = & -e^{-2/\tau}
\end {eqnarray}
For a given time series the parameters $a_i$ can be estimated e.g.  by
the Durbin-Levinson- or Burg-algorithm (Honerkamp 1993). By
statistical testing it is possible to infer whether a model is
compatible with the data.

\begin{figure}
{Fig.2: a) EXOSAT ME X-ray lightcurve of NGC~5506 (Jan.~1986), 
b) Hidden AR[1]-process, estimated with the LSSM fit.}
\end{figure}

A first generalization of AR models are the autoregressive-moving-average 
(ARMA) models that include also past noise terms in the dynamics:
\begin{eqnarray}
    x(t) = \sum_{i=1}^p a_i x(t-i)  +  \sum_{j=1}^q b_j
      \epsilon(t-j) + \epsilon(t)
\end{eqnarray}
Both models, AR and ARMA processes, assume that the time series is
observed without any oberservational noise. In presence of such noise
the parameters $a_i$ will be underestimated and statistical tests will
reject the model even if its order is specified correctly. 

LSSMs generalize the AR and ARMA processes by explicitly modelling
observational noise. Furthermore, LSSMs use the so called Markov
property, which means that the entire information relevant to the
future or for the prediction is contained in the present state. The
variable $x(t)$ that has to be estimated cannot be observed directly
since it is covered by observational noise $\eta(t)$. Following the
Markov property it is possible to regressively predict the values
$x(t)$, though.

The measured observation variables $y(t)$ may not necessarily agree
with the system variables $x(t)$ that provide the best description of
the system dynamics. Thus a LSSM is defined with two equations, the
system or dynamical equation~(\ref{eq:sys}) and the observation
equation~(\ref{eq:obs}).
\begin{eqnarray}
 \vec{x}(t) & = &  {A} \, \vec{x}(t-1) +  \vec{\epsilon}(t) 
 \quad \vec{\epsilon}(t) \sim {\calN{N}} (0, {Q}) \label{eq:sys} \\ 
 y(t) & = &  {C} \, \vec{x}(t) + \eta(t) 
 \quad \eta(t) \sim {\calN{N}}(0,R) \label{eq:obs}
\end{eqnarray}
This definition is a multivariate description, which means that the
AR[p] process is given as a $p$-dimensional AR process of order one,
with a matrix $ {A}$ that determines the dynamics. By combining the
different dimensional terms of the multivariate description the
typical AR[p] (see eq.~\ref{eq:ar}) form can be derived easily. The
observation $y(t)$ is formulated as a linear combination of the random
vectors $\vec{x}(t)$ and $\eta(t)$. The matrix $ {C}$ maps the
unobservable dynamics to the observation. The terms
$\vec{\epsilon}(t)$ and $\eta(t)$ represent the dynamical noise with
covariance matrix $ {Q}$ and the observational noise with variance
$R$, respectively.

The estimation of the parameters in LSSMs is more complicated than for
AR or ARMA processes. There are two conceptually different procedures
available to obtain the maximum likelihood parameters estimates. Both
are iterative and start from some initial values that have to be
specified. The first procedure uses explicit numerical optimization
to maximize the likelihood. The other applies the so called
Expectation-Maximization algorithm. The latter procedure is slower
but numerically more stable than the former and is described in detail
by Honerkamp (1993). Statistical evaluation of a fitted model is
generally based on the prediction errors. The prediction errors are
obtained by a Kalman filter which estimates the unobservable process $
\vec{x}(t) $ (Hamilton 1995). Such a linear filter allows us to arrive
at the variables $\hat{\mbox{$\vec{x}$}}(t)$ (and its prediction
errors), used to describe the system dynamics, starting from a given
LSSM and the given observations $y(t)$ (Brockwell and Davis 1991, Koen
and Lombard 1993).

Multiplying the estimated process $\hat{\mbox{$\vec{x}$}}(t)$ with the
estimated $ {C}$ yields an estimate $ \hat{y}(t) $ of the observed
time series $y(t)$. A necessary condition that the model fits to the
data is that the difference $y(t)-\hat{y}(t)$ represents white
noise, i.e. the time series of prediction errors should be 
uncorrelated. This can for example be judged by a Kolmogorov-Smirnov
test that tests for a flat spectrum of the prediction errors or by the
Portmanteau test using their autocorrelation function. We have used
the first method to quantify the goodness of fit of the tested LSSMs
(see table~1).

Another criterion to judge fitted models is the decrease in the
variance of prediction errors with increasing order of the fitted
models. A knee in this function gives evidence for the correct model
order. Any further increase of the model order will not reduce the
variance significantly. The so called Akaike information criterion
(AIC) formulizes this procedure including the different number of
parameters of the models (Hamilton 1995). Any oscillators and
relaxators which might occur in unnecessarily more complex LSSMs
should be highly damped and can be neglected therefore.

The last method to judge a fitted model is to compare the spectrum that 
results from the fitted parameters with the periodogram of the sample 
time series. The spectrum of a LSSM is given by :
\begin{eqnarray}
      S(\omega) =  {C} ( {1}- {A}e^{-{\rm i} \omega})^{-1}  {Q}
      \left(( {1}- {A}e^{{\rm i} \omega})^{-1}\right)^{\rm T} 
                                                       {C}^{\rm T} + R
      \label{eq:arsp}
\end{eqnarray}
The superscript $T$ denotes transposition. Spectra of AR or ARMA processes
are special cases of equation~(\ref{eq:arsp}). In the simplest case 
of an AR[1] process modelled with a LSSM, the corresponding spectrum 
is given by:
\begin{eqnarray}
S(\omega)_{\rm LSSM\,AR[1]} = \frac{Q}{1 + a_1^2 - 2a_1\cos(\omega)}+R
\end{eqnarray}
This function provides both the flattening at low and the decrease of
power at medium frequencies seen in periodograms (e.g. see Fig.~4).

In a first approach gaps in the observed lightcurve were filled with white
noise with the same mean and rms as the original time series in order to
create a continuous time series. In a second run these gaps were refilled
with the predictions of the Kalman filter plus a white noise realization
with the original lightcurves variance. Generally, gaps in an observed time
series can be handled by the LSSM in a natural way avoiding the filling of
gaps with Poisson noise. The key is again the Kalman filter. The Kalman
filter considers the fact that there are still decaying processes taking
place even if the object is not observed. In each cycle of the iterative
parameter estimation procedure $\vec{x}(t)$ is estimated based on an
internal prediction, corrected by information obtained from the actual data
$y(t)$. In case of gaps no information from $y(t)$ is available and the
internal prediction decays in its intrinsic manner until new information is
given. In the case of the lightcurve of NGC~5506 the resulting parameters
are consistent with those of the first approach due to the high duty cycle
of the original time series.

\section{The EXOSAT observation of NGC~5506} \label{exosat}
As the X-ray lightcurves from EXOSAT are the longest AGN observations
available, we have used the longest individual observation of about
230\,ks of the Seyfert galaxy NGC~5506 for applying the LSSM
(Fig.~2a). The data which have been extracted from the HEASARC EXOSAT
ME archive, are background subtracted and dead time corrected, with a
30 sec time resolution obtained over 1--8\,keV energy range. The
Seyfert galaxy NGC~5506 holds a special place in AGN variability
studies, as it is both bright and one of the most variable AGN. The
chosen lightcurve contains only few gaps providing a duty cyle of
92.4\%. The mean and rms of the lightcurve are 6.87 and 1.55 counts in
30\,s bins.

\begin{table}\centering
\caption{Results of LSSM fits to the EXOSAT NGC 5506 data}
\begin{tabular}[c]{ccccc}
\hline
Model      & $R_{\eta}^a$ & Periods & $\tau^b$ & KS test$^c$ \\
LSSM AR[p] &              & (s)     & (s)      &             \\
\hline
   0 & 1               & -    & -        &   0.0\%   \\
   1 & 0.722           & -    & 4799     &   93.5\%   \\
   2 & 0.701           & -    & 26.1     &   66.8\%   \\
      &                & -    & 5011     &         \\
   3 & 0.510           & -    & 10.6     &   88.2\%   \\
      &                & -    & 18.9     &         \\
      &                & -    & 4798     &         \\
   4 & 0.395           & 236.3 & 71.1    &   92.1\%   \\
      &                & -    & 6.7      &         \\
      &                & -    & 4780     &         \\
\\
\hline
\multicolumn{5}{l}{\small $^a$ Variance of the observational noise}\\
\multicolumn{5}{l}{\small $^b$ Relaxation time}\\
\multicolumn{5}{l}{\small $^c$ Kolmogorov-Smirnov test for white noise}\\
\end{tabular}
\end{table}

We applied LSSMs with different order AR processes. An LSSM using an AR[0]
process corresponds to a pure white noise process without any temporal
correlation and a flat spectrum. The used Kolmogorov-Smirnov test rejects
this model at any level of significance (see table 1). Without loss of
generality, $Q$ is set to unity, the mean and variance are set to 0 and 1,
respectively. We see that the X-ray lightcurve of NGC~5506 can be well
modelled with a LSSM AR[1] model, as the residuals between the estimated
AR[1] process and the measured data are consistent with Gaussian white
noise.  Fig.3 shows the distribution and the corresponding normal quantile
plot of the
fit residuals which both display the Gaussian character of the
observational noise. The standard deviation of the distribution is 0.738
which is in good agreement to the estimated observational variance of 0.722
for the LSSM
AR[1] fit (see table 1). Furthermore, the lightcurve of the estimated AR[1] 
looks very similar to the temporal behavior of the hidden process (Fig.2). The
corresponding dynamical parameter $a_1$ of the LSSM AR[1] fit is 0.9938
which corresponds to a relaxation time of about 4799~s.

\begin{figure}
{Fig.3: a) Distribution and b) normal quantile plot of the residuals of the
LSSM AR[1] fit to the EXOSAT ME NGC 5506 lightcurve (the dotted lines in a)
indicate the mean and rms of the observational noise). A normal quantile
plot arranges the data in increasing order and plot each data value at a
position that corresponds to its ideal position in a normal distribution.
If the data are normally distributed, all points should lie on a straight
line.}
\end{figure}

The LSSM AR[1] gives a good fit to the EXOSAT NGC~5506 data as the variance
of the prediction errors nearly remains constant from model order~1 to~2
and the residuals conforms to white noise. The decrease in the variance for
higher model orders might be due to correlations in the modelled noise,
generated by the switching of the EXOSAT detectors. Since each detector has
its own noise charateristics a regular swapping between background and
source detectors would lead to an alternating observational noise level
(see Section 2). The higher order LSSM AR[p] fits try to model the
resulting correlations with additional but negligible relaxators and damped
oscillators ($\tau \approx {\rm bintime}$, $\tau \ll T_{\rm{tot}}$).

We have used the Durbin-Levinson algorithm (see section~3) to estimate
the parameters of a competing simple AR[p] model (see table~2). As
expected for time series containing observational noise, the
characteristic timescales are underestimated by fitting a simple AR
process and the statistical test rejects the AR[p] model. A test for
white noise residuals fails, which means that there are still
correlations present which cannot be modelled with an AR[p] procces.
We have performed AR[p] fits for model orders up to 10 and we never
found residuals consitent with white noise, indicating that there is
no preferred model order. All occuring relaxators and damped
oscillators are insignificant due to their short relaxation timescales
compared with the bintime of 30\,s. As the observational noise is not
modelled explicitly in AR models, it is included accidentally in the
inherent AR noise term. Thus, any correlation in the observed time series
which can be detected in the LSSM fits, is wiped out and the
higher order AR fits only reveal fast decaying relaxators and
oscillators.

\begin{table}\centering
\caption{Results of AR fits to the EXOSAT NGC 5506 data}
\begin{tabular}[c]{ccccc}
\hline
Model & $Q_{\epsilon}^a$ & Periods & $\tau^b$ & KS test$^c$ \\
AR[p] &                 & (s)     & (s)      &            \\
\hline
    0 & 1               & -    & -        &   0.0\%   \\
    1 & 0.9235          & -    & 23.3     &   0.5\%   \\
    2 & 0.8814          & -    & 55.6     &   0.3\%   \\
      &                 & -    & 29.8     &             \\
    3 & 0.8566          & -    & 97.0     &   0.4\%   \\
      &                 & 197.4 & 40.6    &         \\
    4 & 0.8362          & -    & 153.2    &   0.4\%   \\
      &                 & -    & 51.2     &         \\
      &                 & 127.7 & 55.1    &         \\
\\
\hline
\multicolumn{5}{l}{\small $^a$ Variance of inherent AR noise}\\
\multicolumn{5}{l}{\small $^b$ Relaxation time}\\
\multicolumn{5}{l}{\small $^c$ Kolmogorov Smirnov test for white noise}\\
\end{tabular}
\end{table}

One might expect that the resulting best fit LSSM light curve (Fig.~2b)
might also be produced by just smoothing the original lightcurve. This
assumption is wrong as a smoothing filter would pass long timescales and
suppress all short time variability patterns. Thus all information about
the variations on short timescales would be lost (Brockwell and Davis
1989). The Kalman filter concedes not only the time series values $x(t)$
but also its prediction errors. These errors are much smaller than the
errors of the observed lightcurve $y(t)$. In the case of the NGC~5506
observation (Fig.2) the estimation errors are about 0.18 counts/sec and the
errors of $y(t)$ are about 1.3 counts/sec, respectively. Both lightcurves
in Fig.2 are shown without error bars due to reasons of clarity.

We have used Monte Carlo Simulations to determine the error of the
dynamical parameter $a_1$. Using the distribution of the estimated
parameters of 1000 simulated AR[1] time series with the best fit
results, we found $a_1~=~0.9938 \pm 0.0007$. As the dynamical
parameter is close to unity the corresponding relaxation time error is
high, with $\tau = 4799^{+632}_{-472}$\,s. To prove the quality of the
LSSM results we have fitted a LSSM AR[1] spectrum to the periodogram
data. This fit yields the dynamical parameter $a_1~=~0.9936 \pm
0.0021$ which is consistent with the LSSM AR[1] fit in the time
domain, but the corresponding error is much higher due to the lower
statistical significance of frequency domain fits (see Section 2).

The autocovariance function of the AR[1] process is given by:
\begin{eqnarray}
      {\rm ACF}_{\rm AR[1]}({\Delta}) = \frac {Q} {1 - a_1^2} \,
e^{\log(a_1) \, \Delta} \label{eq:rar}
\end{eqnarray}
which is an exponentially decaying function for stationary ($|a_1| <
1$) time series, very similar to the temporal behavior of the
autocorrelation function of a shot noise model (Papoulis 1991):
\begin{eqnarray}
      {\rm ACF}_{\rm shot\,noise}({\Delta}) = \frac {\lambda \tau}{2}
 \, e^{-\Delta/\tau }
\end{eqnarray}

The variable $\lambda$ denotes the density and $\tau$ is the lifetime
of the shots. This similarity means that an AR[1] process can also be
modelled by a superposition of Poisson distributed decaying shots
(Papoulis 1991). The shot noise model, which has been used as an
alternative to the $1/f^{\alpha}$ model, appears to give a good fit to
the power spectrum of NGC~5506 (Papadakis and Lawrence 1995, Belloni
and Hasinger 1990 and references therein). But instead of all the
shots having the same lifetime, Papadakis and Lawrence (1995) used a
distribution varying as $\tau^{-2}$ between $\tau_1$ and $\tau_2$.
They fixed $\tau_2$ arbitrarily at 12\,000\,s and found that $\tau_1$
is around 300\,s for NGC~5506, much lower than the relaxation time of
about 4800\,s found with the LSSM fit. A possible explanation for this
difference could be the distribution of lifetimes. Since the power law
slope of the shot noise model is constantly $-2$ at medium and high
frequencies, this distribution is necessary to modify the slope and to
maintain a good fit to the spectrum. The advantage of a LSSM is a
variable slope at medium frequencies which depends on the dynamical
parameter (see Fig.~4).

\begin{figure}
{Fig.4: Periodogram of the EXOSAT ME X-ray lightcurve of NGC~5506
(dots) and the spectrum of the best fit LSSM AR[1] model in the time domain
(line) (see Fig.2a). The spectra of the higher order LSSM AR fits differ
less than $2\%$ from the LSSM AR[1] spectrum. The dashed lines display the
$\pm1\sigma$~-~spectra of the corresponding frequency domain fit. The time
domain fit yields $1\sigma$ errors which are more than 3 times smaller
(see text for details).}
\end{figure}

The shot noise model can be regarded as an approximation of an AR[1]
model for values $a_1$ near unity. The mean density of the Poisson
events $\lambda$ then corresponds to the variance $Q$ of the dynamical
noise in the LSSM system equation~(\ref{eq:sys}). Thus $Q$ could be
used to quantify and compare the rate of the accretion shots occuring
in AGNs.

\section {Discussion} 

We obtain a convincing fit to the observed X-ray lightcurve of an AGN using
a LSSM AR[1] process as well in the time and in the frequency domain. The
explicit modelling of observational noise allows to estimate the covered
AR[1] process, indicating that the stochastic process is dominated by a
single relaxation timescale. We show that the general AR[p] model (see
Eq.1) can be restricted to a simple AR[1] process which succeeds in
describing the entire dynamics of the observed AGN X-ray lightcurve.

It has been suggested by McHardy (1988) that the single shots, which are
supposed to be superimposed to build the lightcurve, may arise from
subregions of an overall larger chaotic region which are temporarily lit
up, perhaps by shocks. Since one would expect a non uniform electron
density throughout this region (probably decreasing with distance from the
central engine), the resulting difference in cooling timescales yields the
different decay timescales (Green et al. 1993). As the LSSM predicts that
the stochastic process is dominated by a single relaxator, we presume the
existence of a single cooling timescale or a uniform electron density in
the emission region following the shot noise model (see Sutherland et
al. 1978).

The assumption of an exponentially decaying shot seems to be reasonable as
time-dependent Comptonisation models lead to such a pulse profile. The
scenario for a thermal Comptonisation model (Payne 1980, Liang and Nolan
1983) starts with UV photons which arise as the accretion inflows
inhomogeneities, each producing a single flare when gravitational energy is
set free as radiation. The impulsive emission of the Poisson distributed
delta peaks in a cloud of hot electrons triggers X-ray flares with a
specific pulse profile depending on the seed photon energy, the density,
and the temperature of the electrons.  This impulsive emission is delayed
and broadened in time and spectrally hardened due to repeated Compton
scattering. Some approximate analytic solutions of this process show that
the temporal evolution of the generated X-ray pulse can be described by a
nearly exponentially decaying function (Miyamoto and Kitamoto 1989). The
only difference to the `shots' used above is the (more realistic) non-zero
rise time. Using this model it should be possible to associate the
estimated relaxator timescale $\tau_{\rm AR [1]}$ with the physical
properties of the Comptonisation process.

The presented LSSM can also be used to analyse X-ray variability of
galactic X-ray sources. As both, relaxators and (damped) oscillators
can be estimated, it is possible to use the algorithm to search for
periodocities and QPO phenomena in the lightcurves of X-ray binaries
(see Robinson and Nather 1979, Lewin et al. 1988, van der Klis 1989).

{\it Acknowledgements:} We would like to thank J.D.~Scargle, R.~Staubert,
M.~Maisack, J.~Wilms and K.~Pottschmidt for helpful discussions and
C.~Gantert for writing the code of the LSSM program. Furthermore, we thank
the anonymous referee for constructive comments.

\begin{thebibliography} {999}

Abramowicz A.R., Chen X., Kato, S. et al., 1995, ApJL 438, L37
Begelman M.C., de Kool M., 1991, in Variability in Active Galactic Nuclei, ed. H.R.Miller \& P.J.Wiita (Cambridge: Cambridge Univ.Press), 198
Belloni T., Hasinger G., 1990, A\&A 227, L33
Brockwell P.J., Davis R.A., 1991, Time Series: Theory and Methods, Springer Verlag, 2nd.ed.
Deeter J.E., 1984, ApJ 281, 482-491
Deeter J.E., Boyton P.E., 1982, ApJ 261, 337-350
Fuller W.A., 1996, Introduction to Statistical Time
Series, New York, John Wiley, 2nd.ed.
Grandi P., Tagliaferri G., Giommi P. et al., 1992, ApJ Suppl. Ser., 82, 93
Green A.R., McHardy I.M., Lehto H.J., 1993, MNRAS 265, 664-680
Hamilton J.D., 1995, Time Series Analysis, Princeton University Press
Honerkamp J., 1993, Stochastic Dynamical Systems, VCH Publ. New York, Weinheim 
Koen C., Lombard F., 1993, MNRAS 263, 287
Krolik J.H., 1992, Statistical Challenges in Modern Astronomy, E.Feigelson
and G.J.Babu, eds., Springer Verlag New York, 349
Lawrence A., Watson M.G., Pounds K.A. et al., 1987, Nature 325, 694
Lawrence A., Papadakis P., 1993, ApJ Suppl. 414, 85
Leahy D.A., Darbro W., Elsner R.F., Weisskopf M.C, Sutherland P.G., 1983,
ApJ 266, 160-170
Lehto H.J., 1989, in Proc. 23rd ESLAB Symp. on Two Topics in X-ray Astronomy, Vol.1, ed. J.Hunt \& B.Battrick (ESA SP-296), Noordwijk, 499
Lewin W.H.G., van Paradijs J., van der Klis M., 1988, Space Sci. Review 46, 273
Liang E.P., Nolan P.L., 1983, Space Sci. Review 38, 353
McHardy I., Czerny B., 1987, Nature 325, 696
Mineshige S., Ouchi N.B., Nishimori H. et al., 1994, PASJ 46, 97
Miyamoto S., Kitamoto S., 1989, Nature 342, 773
Netzer H., 1990, Saas-Fee Advanced Course 20 on Active Galactic Nuclei, eds. R.D. Blandford, H.Netzer, L.Woltjer, 57-160
Papadakis I.E., Lawrence A., 1995, MNRAS 272, 161
Papoulis A.P., 1991, Probability, Random Variables and Stochastic 
Processes, New York, McGraw-Hill, 3rd.ed.
Payne D.G., 1980, ApJ 232, 951
Priestley M.B., 1992, Spectral Analysis and Time Series, San Diego, Academic Press
Rees M.J., 1984, Ann.Rev.Astron.Astrophys. 22, 471
Robinson E.L., Nather R.E., 1979, ApJ Suppl. 39, 461
Samorodnitsky G., Taqqu M.S., 1994, Stable Non-Gaussian Random Processes, New York,Chapman and Hall
Scargle J.D., 1981, ApJ Suppl. 45, 1
Sutherland P.G., Weisskopf M.C., Kahn S.M., 1978, ApJ 219, 1029
Tagliaferri G., Bao G., Israel G.L. et al. 1996, ApJ, accepted for publication
van der Klis M., 1989, Ann.Rev.Astron.Astrophys. 27, 517
Wold H.O.A., 1938, A Study in the Analysis of Stationary Time Series,
Uppsala, Almqvist and Wiksell, 2nd.ed.
Yule G., 1927, Phil. Trans. R. Soc. A 226, 267 \\
\end {thebibliography}

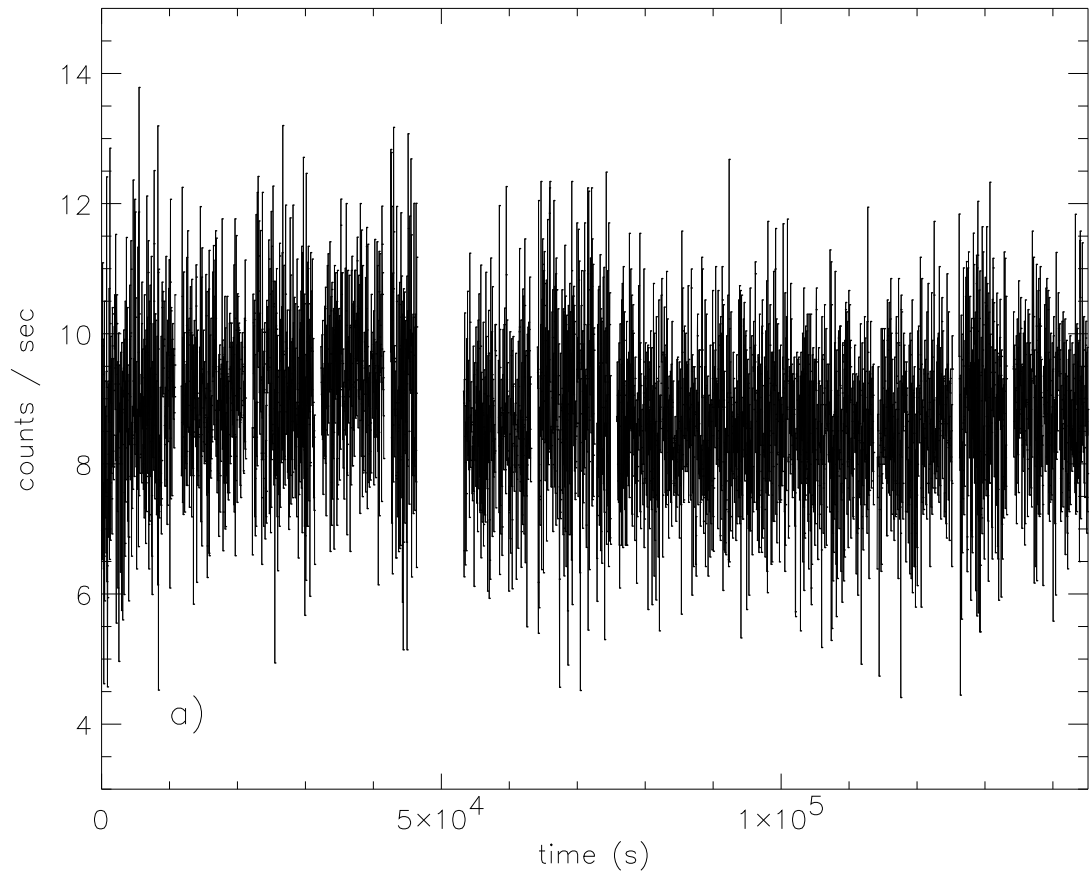
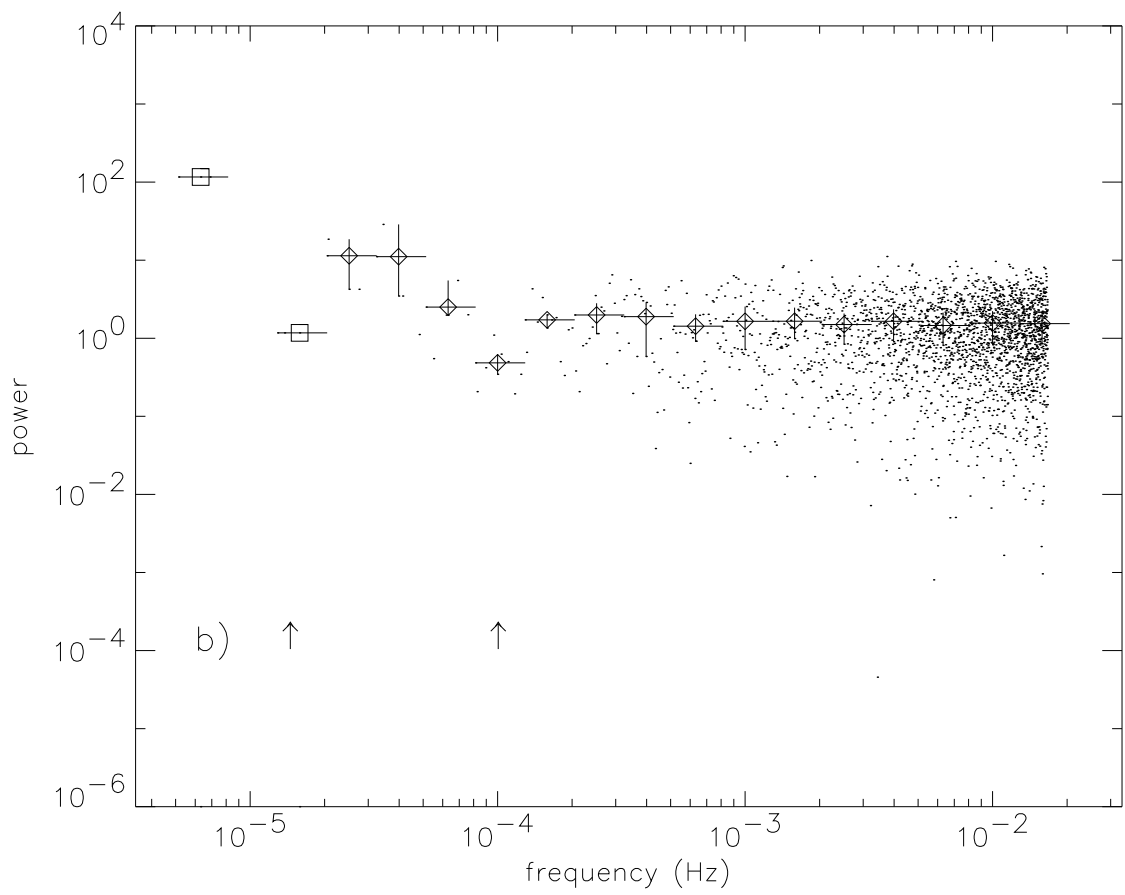
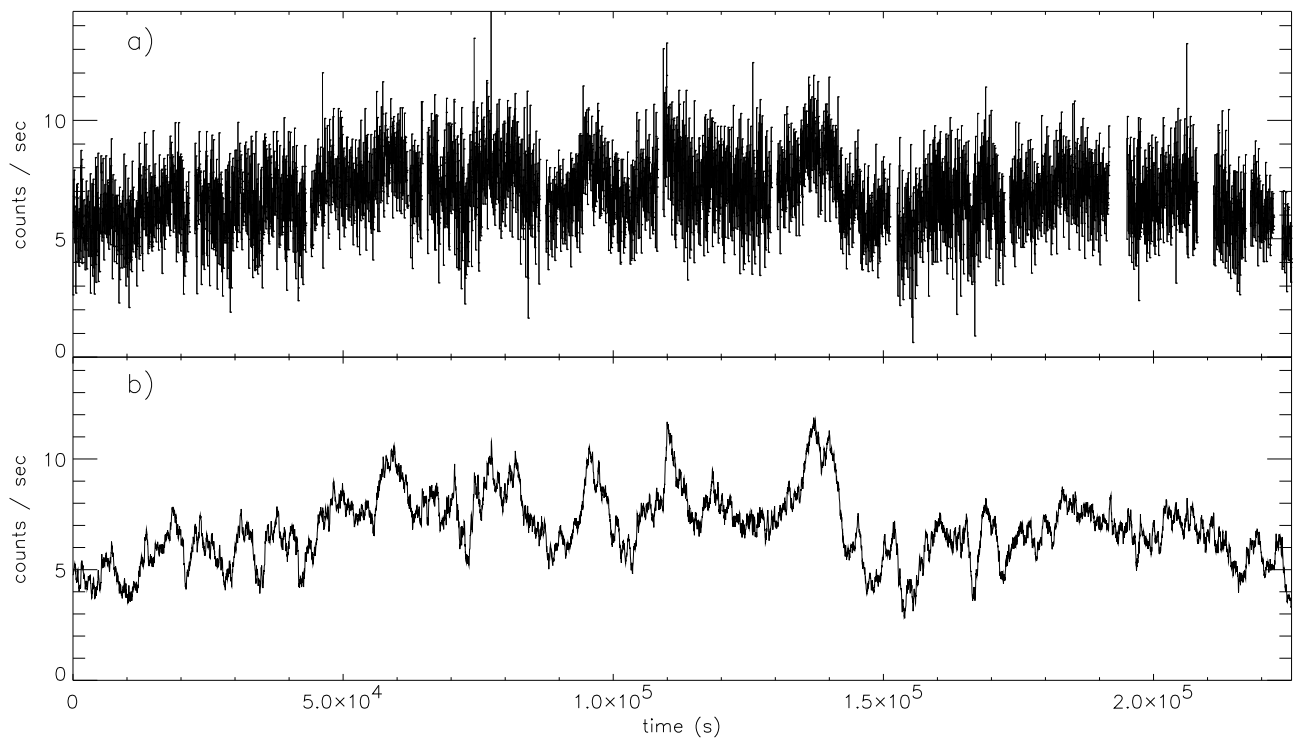
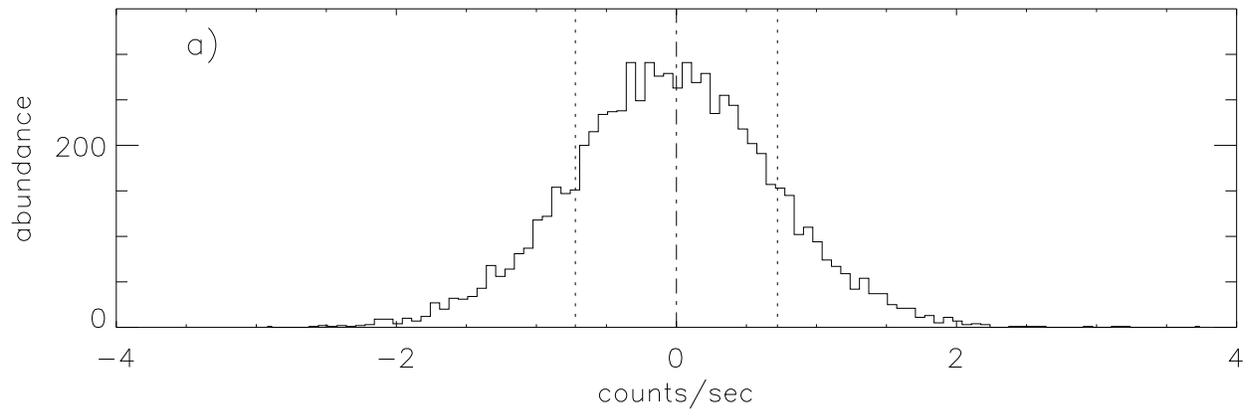
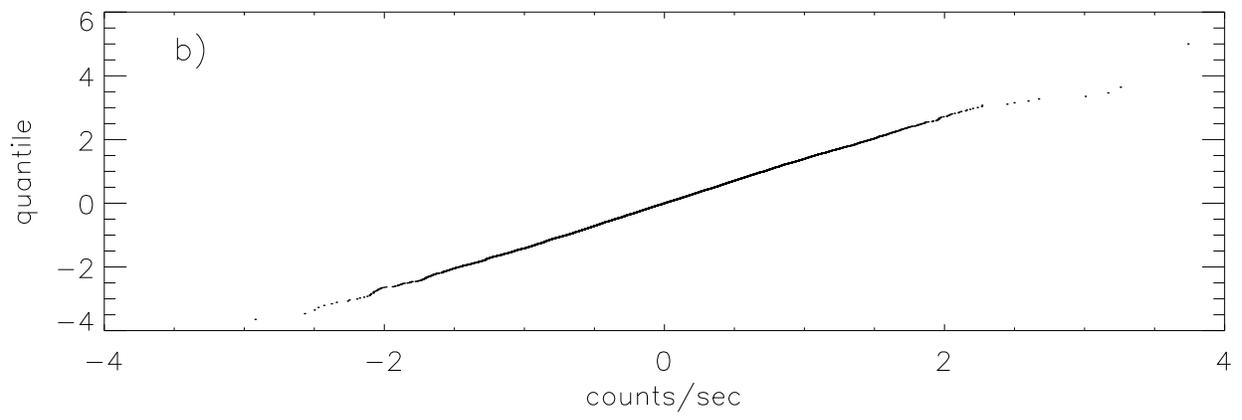
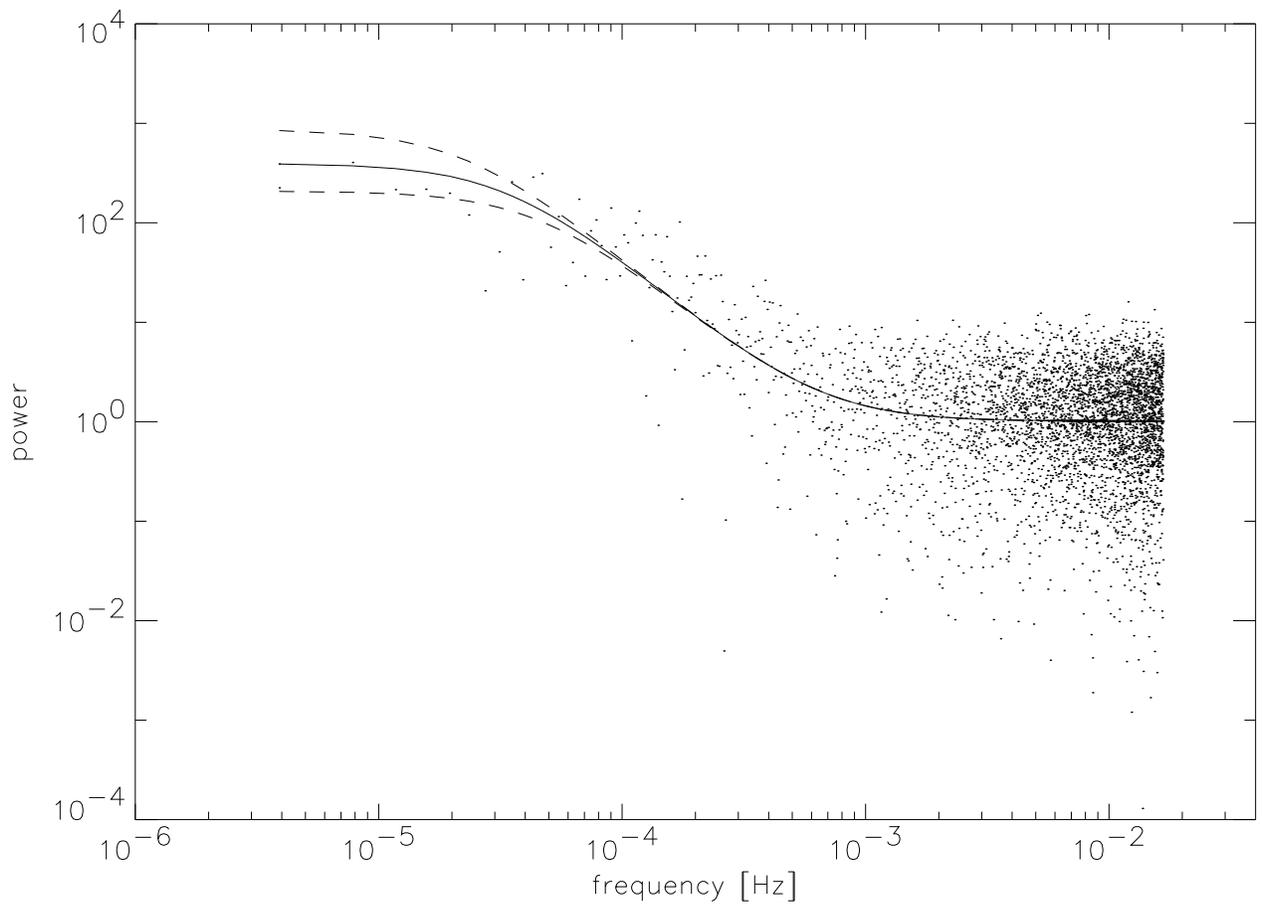
\end{document}